\documentclass[twocolumn]{aastex63}

\usepackage{natbib}
\usepackage{color}
\usepackage{mathtools}
\usepackage{hyperref} 

\def	\cm		{\,{\rm {cm}}}
\def	\K		{\,{\rm K}}

\def	\mum	{\,{\mu \rm{m}}}

\def \bea {\begin{eqnarray}}
\def \ena {\end{eqnarray}}

\def	\bM	{\boldsymbol{M}} 

\def    \bmu    {{\hbox{\boldsym\char'026}}}	

\def	\br	{\boldsymbol{r}}

\def	\cm	{\,{\rm cm}}
\def	\km	{\,{\rm km}}

\def	\max	{\,{\rm max}}

\def	\erg	{\,{\rm erg}}

\def	\gas	{\,{\rm gas}}

\def	\H	{{\rm H}}

\def	\s	{\,{\rm s}}
\def	\sp	{{\rm sp}}

\def	\AU	{\,{\rm au}}

\def \St {{\rm St}}

\def	\rad	{{\rm rad}}
\def	\yr	{{\rm yr}}

\def	\ahat		{\hat{\bf a}}

\def    \Bv     	{\boldsymbol{B}}

\def    \gas     	{{\rm gas}}

\font\mib=cmmib10

\def\bOmega{\hbox{\mib\char"0A}}
\def\bmu{\hbox{\mib\char"16}}

\begin{document}
\shorttitle{Effect of dust magnetic properties on grain growth}
\shortauthors{Hoang and Truong}
\title{Effects of Barnett magnetic dipole-dipole interaction on grain growth and destruction}

\author{Thiem Hoang}
\affiliation{Korea Astronomy and Space Science Institute, Daejeon 34055, Republic of Korea} 
\affiliation{Korea University of Science and Technology, 217 Gajeong-ro, Yuseong-gu, Daejeon, 34113, Republic of Korea}

\author{Bao Truong}
\affiliation{Korea Astronomy and Space Science Institute, Daejeon 34055, Republic of Korea}
\affiliation{Korea University of Science and Technology, 217 Gajeong-ro, Yuseong-gu, Daejeon, 34113, Republic of Korea}

\begin{abstract}
Rapidly spinning magnetic grains can acquire large magnetic dipole moments due to the Barnett effect. Here we study the new effect of Barnett magnetic dipole-dipole interaction on grain-grain collisions and grain growth, assuming grains spun up by radiative torques. We find that the collision rate between grains having embedded iron inclusions can be significantly enhanced due to Barnett magnetic dipole-dipole interaction when grains rotate suprathermally by radiative torques. We discuss the implications of enhanced collision rate for grain growth and destruction in the circumstellar envelope of evolved stars, photodissociation regions, and protostellar environments. Our results first reveal the importance of the dust magnetic properties and the local radiation field on grain growth and destruction.

\end{abstract}

\section{Introduction}

Dust is an essential component of the interstellar medium (ISM) and plays an important role in astrophysical processes, including star formation, gas heating and cooling, planet formation, and surface chemistry \citep{Tielens:2007wo,2011piim.book.....D}. Dust grains grow via the accretion of gas species on the grain surface and grain-grain collisions. Grain destruction processes include grain sputtering, grain shattering by grain-grain collisions, and rotational disruption by centrifugal stress (see \citealt{Hoang.2021} for a review).

Grain-grain collisions play an essential role in the grain growth process in astrophysical environments, including the circumstellar envelope, protostellar environments, and planetary atmospheres. Grain collisions in dense molecular clouds lead to grain growth, which is the first stage of planetesimal formation and planet formation \citep{Wurm.2021}. Grain collisions at high velocities can destroy grains via shattering or evaporation. Therefore, the grain-grain collision rate is crucial for describing grain evolution and planetesimal formation. 

The rate of grain collisions depends on collision cross-section, which is determined by the physical interaction between dust grains. Previous studies on the collision cross-section usually consider hard-sphere collisions for neutral grains \citep{1997ApJ...480..647D} or Coulomb interaction for charged grains \citep{Okuzumi:2009p4215}. However, astrophysical dust grains can have magnetic dipole moments due to the existence of paramagnetic material within the dust.

Indeed, dust grains containing embedded iron atoms (e.g., silicate or composite dust grains) can acquire a magnetic moment through three processes, including (1) spontaneous magnetization for ferromagnetic material, (2) induced magnetization by an ambient magnetic field for paramagnetic/superparamagnetic material, (3) and grain rotation (Barnett effect, \citealt{Barnett:1915p6353}). Note that spinning charged grains can also acquire magnetic moment \citep{Martin:1976p7959}, but it is less efficient than the Barnett effect (see \citealt{2016ApJ...816L..32H}).

The dipole-dipole interaction of the grain's magnetic moments produced by the first two magnetization processes has been previously studied \citep{1994Icar..107..155N,2016ApJ...826..152H}. An experimental study by \cite{1994Icar..107..155N} showed that grain growth is efficient for metallic (iron particles). The authors explained the effect based on the attraction of grains by spontaneous magnetic dipole-dipole interaction, which is valid for single-domain ferromagnetic nanoparticles of size $a\sim 10$ nm only \citep{1995Icar..117..431N}. \cite{2016ApJ...826..152H} considered the dipole-dipole interaction of induced magnetic moments for ferromagnetic grains in protoplanetary disks. This mechanism is more efficient in stronger magnetic fields (closer to the star) and was appealed to explain the formation of iron-rich Mercury \citep{Kruss.2020,McDonough.2021}.

In this paper, we study the effect of dipole-dipole interaction where the magnetic dipole moment is induced by the Barnett effect (henceforth Barnett magnetic moment). The Barnett magnetic moment of a magnetic grain of susceptibility $\chi(0)$ rotating at $\Omega$ is proportional to the angular velocity as $\mu_{\rm Bar}\propto \chi(0)\Omega$ \citep{1976Ap&SS..43..291D}. Therefore, faster-rotating grains acquire larger magnetic moments. 

Astrophysical grains tend to rotate suprathermally due to surface processes \citep{1979ApJ...231..404P}, radiative torques \citep{1976Ap&SS..43..291D,1996ApJ...470..551D}, and mechanical torques \citep{2007ApJ...669L..77L,2018ApJ...852..129H}. \cite{2007MNRAS.378..910L} introduced an Analytical Model (AMO) of radiative torques (RATs), which is based on a helical grain consisting of an oblate spheroid and a weightless mirror. The AMO is shown to reproduce the basic properties of RATs obtained from numerical calculations for realistically irregular grain shapes \citep{2007MNRAS.378..910L,Hoang:2008gb,Herranen.2021}, and enables us to make quantitative predictions for various conditions \citep{2014MNRAS.438..680H} and dust compositions \citep{2008ApJ...676L..25L,2009ApJ...695.1457H,2016ApJ...831..159H}. 

The rest of our paper is structured as follows. In Section \ref{sec:dipole} we describe the grain magnetic dipole of spinning magnetic grains induced by the Barnett effect and the effect of dipole-dipole interaction on the grain-grain collision cross-section. In Section \ref{sec:RAT} we quantify the Barnett dipole-dipole interaction for grains spun-up by RATs. The implications of our study for grain growth and destruction in different environments are discussed in Section \ref{sec:discuss}. Our main findings are summarized in Section \ref{sec:summary}.

\section{Barnett Magnetic Dipole-Dipole Interaction}\label{sec:dipole}
\subsection{Grain rotation}
Grains in the ISM rotate due to various interaction processes with ambient gas, radiation, and cosmic rays. In the absence of the latter interactions, grains can achieve thermal energy equilibrium. For an oblate spheroidal grain shape of the semi-major axis of length $a$, the thermal angular velocity of the grain rotation along the symmetry axis in the gas of temperature $T_{\gas}$ is $\Omega_{\rm T}=(kT_{\rm gas}/I_{\|})^{1/2}\simeq 1.66\times 10^{5}\hat{\rho}^{-1/2}T_{2}^{1/2}s^{-1/2}a_{-5}^{-5/2}\rm rad\s^{-1}$ with $T_{2}=T_{\gas}/100\K$ and $s<1$ the axial ratio. The effective size of the spheroidal grain $a_{\rm eff}=sa^{1/3}$. Dust grains can be spun-up to an angular momentum greater than its thermal value (i.e., {\it suprathermal} rotation) by surface processes \citep{1979ApJ...231..404P}, radiative torques due to interaction with radiation and mechanical torques due to gas flow. To describe the grain suprathermal rotation, we introduce a dimensionless parameter, $\St=\Omega/\Omega_{T}$, which is referred to as the {\it suprathermal rotation number}. 

\subsection{Magnetic susceptibility}
\label{sec:mag_suscep}
\subsubsection{Composite grains with embedded iron inclusions}
We consider the composite model of dust grains which contain embedded iron clusters. Let $N_{\rm cl}$ be the number of iron atoms per cluster and $\phi_{\rm sp}$ be the volume filling factor of iron clusters. In thermal equilibrium, the average magnetic moment of the ensemble of magnetic inclusions can be described by the Langevin function with argument $m H/kT_{d}$, where $m= N_{\rm cl}\mu_{0}$ with $\mu_{0}$ being the magnetic moment per Fe atom in iron clusters is the total magnetic moment of the cluster, and $H$ is the applied magnetic field (see e.g., \citealt{Jones:1967p2924}). 

The composite grain exhibits superparamagnetic behavior, which has the magnetic susceptibility given by the Curie law
\bea
\chi_{\sp}(0)=\frac{n_{\rm cl}m^{2}}{3kT_{d}},\label{eq:chisp}
\ena
where $n_{\rm cl}$ is the number of iron clusters per unit volume. Following \cite{2016ApJ...831..159H}, the zero-frequency superparamagnetic susceptibility can be written as 
\bea
\chi_{\rm sp}(0)\approx 0.052N_{\rm cl}\phi_{\rm sp}\hat{p}^{2}\left(\frac{10\K}{T_{d}}\right),\label{eq:chi_sp}
\ena
where $\phi_{\rm sp}$ is the volume filling factor of iron clusters, and $\hat{p}=p/5.5$ with $p= \mu_{0}/\mu_{B}$ with $\mu_{0}$ being the atomic magnetic moment and $\mu_{B}=e\hbar/2m_{e}c$ Bohr magneton. Above, $N_{\rm cl}$ spans from $\sim 20$ to $10^{5}$ (\citealt{Jones:1967p2924}), $\phi_{\rm sp}\sim 0.3$ if $100\% $ of Fe abundance present in iron clusters.  

\subsubsection{Ferromagnetic grains}
Ferromagnetic grains have anisotropic ferromagnetic magnetization, and only the magnetic component $\Bv$ perpendicular to the magnetic moment $\bM_s$ contributes to the magnetic susceptibility (i.e., $\chi_{\parallel}(0) = 0$ and $\chi_{\perp}(0)$; see \citealt{1999ApJ...512..740D}). Thus, the effective magnetic susceptibility at zero-frequency is calculated as
\bea
\label{chi_ferro}
\chi_{\rm ferro}(0) = \frac{2\phi\chi_{\perp}(0)/3}{1 + (4\pi/3)\chi_{\perp}(0)[1 - 2\phi/3]},\label{eq:chi_ferro}
\ena
where $\phi$ is the volume filling factor of single-domain ferromagnetic inclusions. We consider 1\% of single-domain ferromagnetic particles randomly distributed in large grains with $\chi_{\perp}(0) = 3.3$, therefore, $\chi_{\rm ferro}(0)\sim 0.016$. 

\subsection{Barnett magnetic moment}
A magnetic grain of zero-frequency susceptibility, $\chi(0)$, rotating with an angular velocity $\bOmega$ becomes magnetized via the Barnett effect \citep{Barnett:1915p6353}. According to the Barnett effect, atomic electrons within a rotating grain of angular velocity $\Omega$ are subject to an equivalent magnetic field of strength given by
\bea
B_{\rm Bar}&=&\frac{\Omega}{\gamma_{e}}=\frac{2m_{e}}{eg_{e}}\Omega\nonumber\\
&\simeq & 9.4\hat{\rho}^{-1/2}T_{2}^{1/2}s^{-1/2}a_{-5}^{-5/2} \left(\frac{\Omega}{\Omega_{T}}\right)~{\rm mG},\label{eq:Beq}
\ena
$\gamma_{e}=-g_{e}\mu_{B}/\hbar$ the electron gyromagnetic ratio where $g_{e}\approx 2$. The Barnett magnetic field is much larger than the magnetic field of the diffuse ISM with $B\sim 10\,\rm\mu G$  (see \citealt{Crutcher:2010p318}).

The acquired magnetic moment of the grain of volume $V=4\pi a_{\rm eff}^{3}/3$ is then given by
\bea
\mu_{\rm Bar}&=& \chi(0)V{B}_{\rm Bar}=\frac{\chi(0)V}{\gamma_{e}} \Omega,\nonumber\\
&=& \frac{4\pi a_{\rm eff}^{3}\chi(0)}{3\gamma_{e}} \Omega\label{eq:muBar}
\ena
where $\chi(0)=\chi_{\rm sp}(0)$ for superparamagnetic grains.

The magnitude of the Barnett magnetic moment for SPM grains is
\bea
\mu_{\rm Bar, SPM}\simeq 9.2\times 10^{-17}T_{2}^{1/2}a_{-5}^{1/2} St\left(\frac{N_{cl,4}\phi_{\rm sp,-2}}{T_{d,1}}\right)~{\rm esu},~~~\label{eq:muBar_mag}
\ena
where $\phi_{\rm sp,-2}=\phi_{\rm sp}/10^{-2}$ where the normalization factor of $10^{-2}$ corresponds to $3\%$ of iron abundance embedded in the dust in the form of iron clusters (see \citealt{2016ApJ...831..159H}).

The Barnett effect can also magnetize grains with single-domain ferromagnetic inclusions. By substituting with Equation \ref{eq:chi_ferro}, the Barnett magnetic moment for large ferromagnetic grains is calculated as
\bea
\mu_{\rm Bar, ferro} \simeq 2 \times 10^{-19} a_{-5}^{1/2}T_{1}^{1/2} St~ \rm esu.\label{eq:muBar_fer}
\ena

\subsection{Spontaneous and induced magnetic moment}
Magnetic grains can acquire a magnetic moment by the ambient magnetic field. The induced magnetic moment for SPM grains is given by
\bea
\mu_{\rm ind, SPM}&=&\chi(0)VB\nonumber\\
&\simeq & 2.1\times 10^{-19}N_{cl,4}\phi_{\rm sp,-2}a_{-5}^{3}B_{1}T_{1}^{-1}~{\rm esu}.\label{eq:mu_ind}
\ena

Ferromagnetic grains have a spontaneous magnetic moment. The magnetic moment is given by
\bea
\mu_{\rm spon,ferro}&=&M_{\rm Fe}(0)V_{\rm gr}\nonumber\\
&\simeq & 7.3\times 10^{-12} a_{-5}^{3}~{\rm esu},\label{eq:muFe}
\ena
where $M_{\rm Fe}=22000/(4\pi)$ G is the spontaneous magnetization of single-domain ferromagnetic material (see \citealt{2016ApJ...821...91H}).


Comparing Equations (\ref{eq:mu_ind}) and (\ref{eq:muFe}) to Equation (\ref{eq:muBar_mag}) and (\ref{eq:muBar_fer}), one can see that the induced magnetic moment is weakest for the typical ISM magnetic field of $10\,\rm\mu$G, whereas the Barnett moment and spontaneous moment are stronger. Therefore, we disregard the induced magnetic moment and consider the last two processes.

\subsection{Magnetic dipole-dipole interaction and enhanced collision rate}
\subsubsection{Dipole-Dipole interaction potential}
Let $\mu$ and $\mu'$ be the magnetic moments of two grains. A magnetic moment produces a magnetic potential at a large distance $r$ from the dipole, which is given by (\citealt{Landau:1984ui})
\bea
\phi(r)=\frac{\bmu.{\br}}{r^{3}},
\ena
and the magnetic field
\bea
{\Bv}(r)=-\nabla \phi(r).
\ena

The energy potential due to dipole-dipole interaction of two grains of dipole moments $\bmu$ and $\bmu'$, separated by a distance $r$, is given by (see e.g., \citealt{1995Icar..117..431N})

\bea
U_{m}&=&-\bmu'.{\Bv}(r)=(\bmu'.\nabla)\phi(r)\nonumber\\
&=&\frac{\bmu.\bmu'}{r^{3}}-3\frac{(\bmu.r)(\bmu'.r)}{r^{5}}.\label{eq:Um_general}
\ena

For the case of two grains aligned with $\bOmega$ along the magnetic field, then, $\bmu\| \bmu'$. Therefore, the interaction energy becomes
\bea
U_{m}=-\frac{2\mu.\mu'}{r^{3}}.\label{eq:Um_para}
\ena

Using $\mu_{\rm Bar}$ from Equation (\ref{eq:muBar}), one obtains the Barnett dipole-dipole potential energy,
\bea
U_{m}=-0.05 N_{cl,4}^{2}\phi_{\rm sp,-1}^{2}T_{1} \left( \frac{St^{2}}{a_{-5}^{2}}\right)~eV,\label{eq:Um_Bar}
\ena
which implies $U_{m}\sim 5$ eV for $St\sim 10$ and $50$ eV for $St\sim 100$. Therefore, the dipole-dipole potential is significant for grains with suprathermal rotation.

\subsubsection{Enhanced grain collision rate by Barnett dipole-dipole interaction}
Let $b$ be the impact factor of the collisions, $v_{\rm gg}$ is the initial relative velocity of grains at large distance of $r\gg a$. The cross-section of grain-grain collision is defined by the maximum impact parameter, $b_{\max}$, that two grains still collide.

The angular momentum conservation yields the initial value equal to the value at the collision at distance $r_{c}$,
\bea
bmv_{\rm gg}= r_{c}mv_{c}, v_{c}=\frac{bv_{\rm gg}}{r_{c}}.\label{eq:momentum}
\ena

The total energy of the two-body interaction system is then given by
\bea
E_{\rm eff}&=&-\frac{2\mu.\mu'}{r_{c}^{3}}+\frac{mv_{c}^{2}}{2}=-\frac{2\mu.\mu'}{r_{c}^{3}}+\frac{m_{\rm gr}b^{2}v_{\rm gg}^{2}}{2r_{c}^{2}}\nonumber\\
&=&\frac{m_{\rm gr}v_{\rm gg}^{2}}{2}.\label{eq:Etot}
\ena

\begin{figure}
    \centering
    \includegraphics[width = 0.5\textwidth]{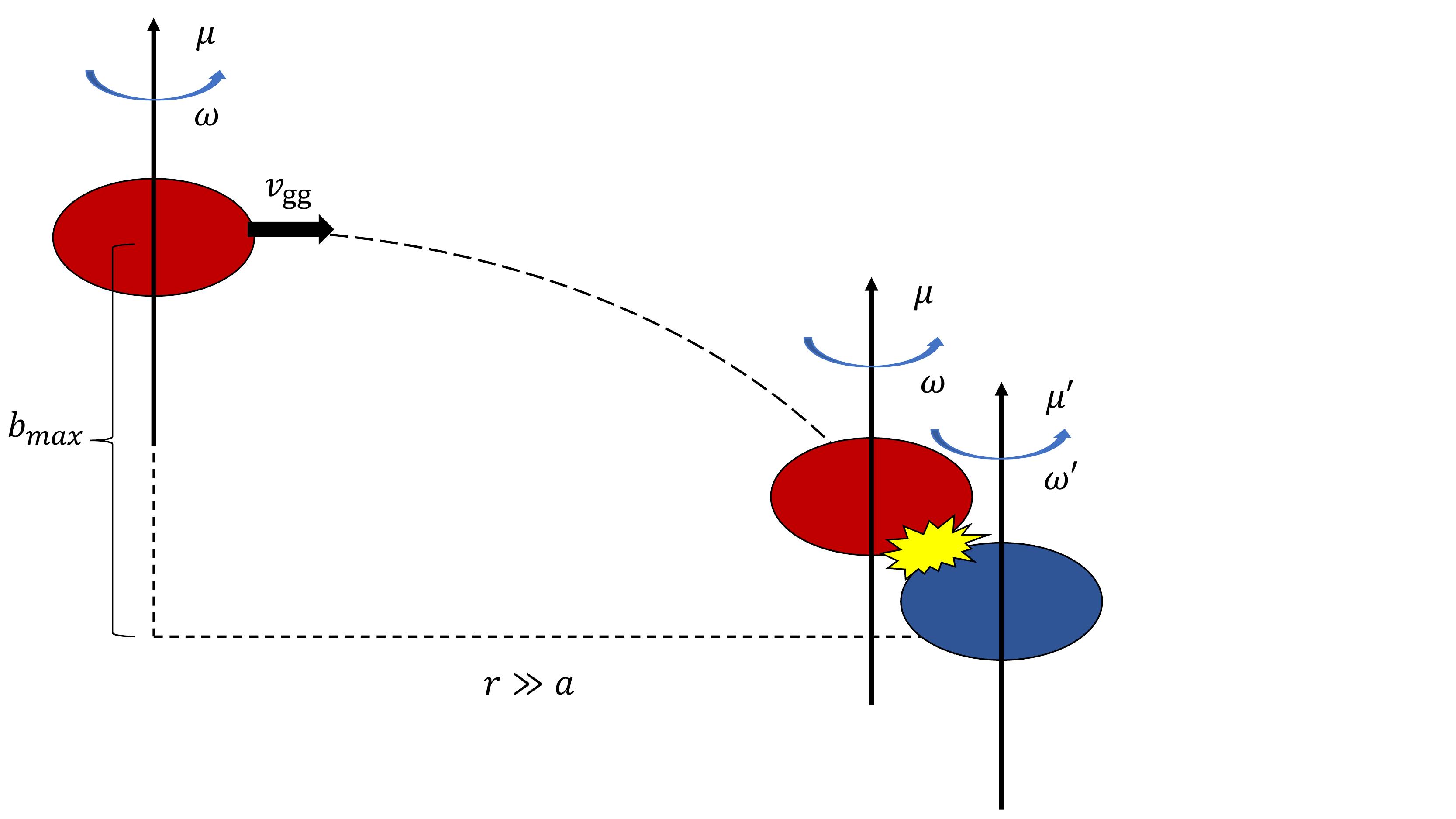}
    \caption{Schematic illustration of the magnetic dipole-dipole interaction between two grains of equal size ($a$) with magnetic moments $\mu$ and $\mu'$ induced by the Barnett effect. The projectile grain is moving with an initial velocity $v_{\rm gg}$ relative to the target grain at rest with the impact parameter $b_{\rm max}$.}
    \label{fig:Dipole_interaction}
\end{figure}

The collision between two grains of equal size occur when $r_{c}=a$. Then, plugging $r_{c}=a$ into Equation \ref{eq:Etot}, one obtains the maximum impact factor $b_{\max}$ as given by
\bea
b^{2}_{\max}&=&a^{2}\left(1+4\frac{\mu\mu'}{m_{\rm gr}a^{3}v_{\rm gg}^{2}}\right),\nonumber\\
&=&a^{2}\left(1+4\frac{3\mu\mu'}{4\pi \rho a^{6}v_{\rm gg}^{2}}\right),\label{eq:bmax2}
\ena 
where $m_{\rm gr}$ is the reduced mass of two grains, which is $4\pi \rho a^{3}/3$ for grains of equal size. The process of grain collision by Barnett dipole-dipole interaction is illustrated in Figure \ref{fig:Dipole_interaction}. \footnote{Our formula differs by a factor 4 in \cite{1995Icar..117..431N}.}

The rate of grain-grain collisions of two equal size grains is then given by
\bea
R_{\rm coll}=n_{\rm gr}v_{\rm gg}\pi b^{2}_{\rm max},\label{eq:Rcoll}
\ena
where $n_{\rm gr}$ is the number density of grain size $a$. 

Therefore, the enhancement in collision rate by Barnett dipole-dipole interaction can be described by $R_{\rm coll}/(n_{\rm gr}v_{\rm gg}\sigma_{\rm geo})=b_{\max}^{2}/a^{2}$ where $\sigma_{\rm geo}=\pi a^{2}$ is the grain geometrical cross-section.

\section{Applications for suprathermal grains by RATs}\label{sec:RAT}

\subsection{Suprathermal rotation by RATs}

Let $\gamma$ and $\bar{\lambda}$ be the anisotropy degree and the mean wavelength of the radiation field. Following \cite{Hoang.2021z5d} (see also \citealt{Hoang.2021}), an irregular grain of effective size $a_{\rm eff}$ subject to a luminous radiation field can be spun up by RATs to a maximum angular velocity given by
\bea
\Omega_{\rm RAT}&=& \frac{3\gamma u_{\rm rad}a_{\rm eff}\bar{\lambda}^{-2}}{1.6n_{\rm H}\sqrt{2\pi m_{\rm H}kT_{\rm gas}}}\left(\frac{1}{1+F_{\rm IR}}\right)\nonumber\\
&\simeq &9.4\times 10^{8} s^{1/3}a_{-5}\left(\frac{\bar{\lambda}}{1.2\mum}\right)^{-2}
\left(\frac{\gamma U}{n_{1}T_{1}^{1/2}}\right)\nonumber\\
&&\times\left(\frac{1}{1+F_{\rm IR}}\right)\rad\s^{-1},\label{eq:omega_RAT1}
\ena
for $a_{\rm eff}\lesssim a_{\rm trans}$ with $a_{\rm trans}=\bar{\lambda}/2.5$ being the transition size at which the average RAT efficiency changes the slope. 

For large grains with $a_{\rm eff}> a_{\rm trans}$, one has
\bea
\Omega_{\rm RAT}&=&\frac{1.5\gamma u_{\rm rad}\bar{\lambda}a_{\rm eff}^{-2}}{12n_{\rm H}\sqrt{2\pi m_{\rm H}kT_{\rm gas}}}\left(\frac{1}{1+F_{\rm IR}}\right)\nonumber\\
&\simeq& 8.1\times 10^{10}s^{-2/3}a_{-5}^{-2}\left(\frac{\bar{\lambda}}{1.2\mum}\right) \left(\frac{\gamma U}{n_{1}T_{1}^{1/2}}\right)\nonumber\\
&&\times\left(\frac{1}{1+F_{\rm IR}}\right)\rad\s^{-1}.\label{eq:omega_RAT2}
\ena

The suprathermal rotation number for the grain spin-up by RATs is then,
\bea
\St_{\rm RAT}&=&\frac{\Omega_{\rm RAT}}{\Omega_{T}}\nonumber\\
&\simeq &1.8\times 10^{4} \hat{\rho}^{1/2}s^{5/6}a_{-5}^{7/2}\left(\frac{\bar{\lambda}}{1.2\mum}\right)^{-2}
\left(\frac{\gamma U}{n_{1}T_{1}}\right)\nonumber\\
&\times&\left(\frac{1}{1+F_{\rm IR}}\right),\label{eq:S_RAT1}
\ena
and
\bea
\St_{\rm RAT}&\simeq& 2.1\times 10^{6}\hat{\rho}^{1/2}s^{-1/6}a_{-5}^{1/2}\left(\frac{\bar{\lambda}}{1.2\mum}\right) \left(\frac{\gamma U}{n_{1}T_{1}}\right)\nonumber\\
&\times&\left(\frac{1}{1+F_{\rm IR}}\right),\label{eq:S_RAT2}
\ena

\subsection{Grain alignment by magnetically enhanced RAT (MRAT)}
Dust grains in the ISM tend to have efficient internal alignment of the grain axis of maximum inertia with its angular momentum due to various internal relaxation processes, including Barnett relaxation, inelastic relaxation. 

For external alignment of the grain angular momentum with the magnetic field, \citep{2016ApJ...831..159H} demonstrated that grains with iron inclusions can have perfect external alignment due to the joint effect of RATs and enhanced magnetic relaxation (cf. \citealt{1951ApJ...114..206D}), which is called magnetically enhanced RAT (i.e., MRAT) alignment. Therefore, magnetic grains are aligned with the magnetic field with $\ahat_{1}\|\bOmega\|\Bv$. As the result, then Barnett magnetic moments of grains are parallel to each other. The perfect alignment of grains with the magnetic field induces the magnetic dipole-dipole interaction to occur at the maximum level because the angle between magnetic dipoles $\theta=0$. 

One can easily check the effect of magnetic relaxation by comparing their timescales with the grain randomization by gas collisions. For grains with iron inclusions, the characteristic time of superparamagnetic relaxation is given by (\citealt{2016ApJ...831..159H}) 
\bea
\tau_{\rm mag,sp} &=& \frac{I_{\|}}{VK_{\rm sp}(\Omega)B^{2}}=\frac{2\rho a^{2}}{5K_{\rm sp}(\Omega)B^{2}},\nonumber\\
&\simeq &
0.15\frac{\hat{\rho}a_{-5}^{2}}{N_{\rm cl}\phi_{\rm sp}\hat{p}^{2}B_{3}^{2}}\frac{T_{d,1}}{k_{\rm sp}(\Omega)}
 \yr,\label{eq:tau_DG}~~~
\ena
where $k_{\rm sp}$ is the function of the rotation frequency (see \citealt{Hoang.20220o}).

The efficiency of magnetic relaxation is described by the ratio of the magnetic relaxation rate to the randomization rate of grain orientation by gas collisions (see \citealt{2016ApJ...831..159H}),
\bea
\delta_{\rm mag,sp}&=&\frac{\tau_{\rm mag,sp}^{-1}}{\tau_{\rm gas}^{-1}}\nonumber\\
&= &5.6\times 10^{3}a_{-5}^{-1}\frac{N_{\rm cl,4}\phi_{\rm sp,-2}\hat{p}^{2}B_{3}^{2}}{\hat{\rho} n_{5}T_{1}^{1/2}}\frac{k_{\rm sp}(\Omega)}{T_{d,1}},\label{eq:delta_m}~~~~
\ena
which implies $\delta_{\rm mag,sp}=56, 5600$ for $N_{\rm cl,4}=0.01, 1$, respectively, assuming $a=0.1\mum$, $n_{5}=1$ and $B=10^{3}\,\rm\mu G$. The efficiency of MRAT alignment depends on $\delta_{\rm mag}$ and become perfect for $\delta_{\rm mag,sp}> 10$ \citep{2016ApJ...831..159H}. 

\subsection{Numerical Results}
For our numerical calculations in this section, we assume that dust grains move randomly with Brownian motion relative to the ambient gas. The grain thermal speed is defined by
\bea
v_{\rm th}=\left(\frac{2kT_{\gas}}{m_{\rm gr}}\right)^{1/2}\simeq 1.5 T_{2}^{1/2}a_{-5}^{-3/2} \cm \s^{-1},\label{eq:vTd}
\ena
which implies $v_{\rm th}=1.5$ and $4.6\cm\s^{-1}$ for the gas of $T_{\gas}=100$ and $1000\K$, respectively. Larger grains have lower thermal speed due to their larger mass. 

For the collisions of two equal-size grains of size $a<a_{\rm trans}$ spun-up by RATs, from Equations (\ref{eq:muBar_mag}) and (\ref{eq:bmax2}), one obtains the enhancement in the grain collision rate by Barnett dipole-dipole interaction as follows
\bea
\frac{b^{2}_{\max}}{a^{2}}\simeq 1+12\frac{a_{-5}^{2}N_{\rm cl,4}^{2}\phi_{\rm sp,-2}^{2}}{v_{-2}^{2}}\left(\frac{\gamma U}{n_{1}T_{1}}\right)^{2}\left(\frac{1.2\mum}{\bar{\lambda}}\right)^{4},~~~~~\label{eq:ba_RAT_SPM1}
\ena
where $v_{-2}=v/10^{-2}\km\s^{-1}$. The equation implies that the cross-section can be enhanced by a factor of 10 for $a_{-5}=1$ or $a=0.1\mum$, assuming the normalized parameters.

For large grains of $a>a_{\rm trans}$, one obtains
\bea
\frac{b^{2}_{\max}}{a^{2}}\simeq 1+4.4\times 10^{4}\frac{N_{\rm cl,4}^{2}\phi_{\rm sp,-2}^{2}}{v_{-2}^{2}a_{-5}^{4}}\left(\frac{\gamma U}{n_{1}T_{1}}\right)^{2}\left(\frac{\bar{\lambda}}{1.2\mum}\right)^{2},~~~~~\label{eq:ba_RAT_SPM2}
\ena
which decreases with increasing the grain size.

Equations (\ref{eq:ba_RAT_SPM1}) and (\ref{eq:ba_RAT_SPM2}) reveal that the enhancement in the grain collision rate by Barnett dipole-dipole effect depends on dust magnetic properties ($\phi_{\rm sp},N_{\rm cl}$), grain sizes, the radiation field, and gas properties, $\gamma U/n_{1}T_{1}$, and the grain relative velocity. For the typical values, one obtains $(b_{\max}/a)^{2}\sim 12$ for $a=a_{\rm trans}=0.5\mum$ for $\bar{\lambda}=1.2\mum$.

\begin{figure*}
    \centering
    \includegraphics[width = 1\textwidth]{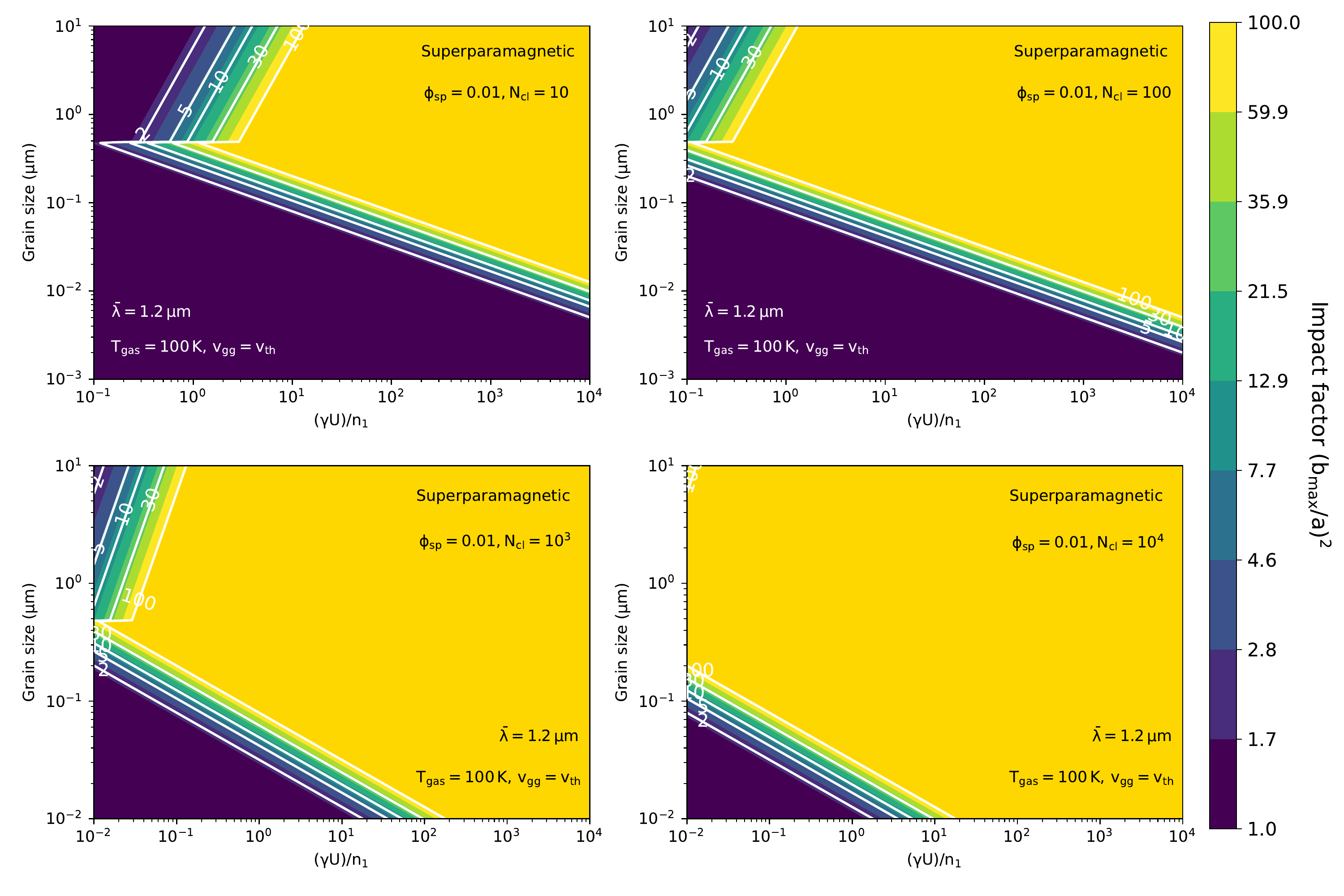}
    \caption{Enhanced collision rate of SPM grains by Barnett dipole-dipole interaction assuming grains having Brownian motion in the gas of temperature $T_{\rm gas}=100\K$, calculated for different iron inclusions, $N_{\rm cl}$. The Barnett dipole-dipole interaction is more effective (i.e., $(b_{\rm max}/a)^2 > 100$) for larger grain sizes due to lower thermal speed.}
    \label{fig:bmax_a_SPM_vther_T100}
\end{figure*}

Figure \ref{fig:bmax_a_SPM_vther_T100} shows the enhancement in grain collision rate due to Barnett magnetic dipole-dipole interaction as a function of the grain sizes for different radiation fields and gas density, denoted by $\gamma U/n_{1}$, assuming grains moving at thermal speeds $v_{\rm gg} = v_{\rm th}$. Different levels of iron inclusions are considered with $N_{\rm cl} = 10 - 10^4$. 

One can see that, under a luminous radiation field (i.e., $\gamma U/n_{1} > 1$), the impact of radiative torques is more significant (i.e., high $\Omega_{\rm RAT}$), leading to the enhancement of the magnetic moment by Barnett effect (see Equation \ref{eq:muBar}). The enhancement links to the level of iron inclusions and increases with $N_{\rm cl}$. Even with a small iron cluster of $N_{\rm cl}=10$, the Barnett dipole-dipole interaction is still important. The parameter space for enhancement is broadened for higher radiation fields ($\gamma U/n_1$) and large grain sizes. 
 
As an example, for SPM grains $a = 0.1\,\rm \mu m$ with $N_{\rm cl} = 10$ (upper left panel of Figure \ref{fig:bmax_a_SPM_vther_T100}), the collision cross-section $(b_{\max}/a)^{2}$ could increase up to 10 when $\gamma U/n_{1} \sim 10$. The impact parameter is higher with $(b_{\max}/a)^{2} > 100$ for larger grains $a > 1\,\rm \mu m$ as a result of lower thermal speed (see Equation \ref{eq:vTd}). In contrast, in the environments with higher gas density or weak radiation field (i.e., $\gamma U/n_{1} < 1$), the grain rotation is affected by the damping by gas collisions, which slows down the grain rotation. Consequently, the collision rate is not enhanced by the Barnett dipole-dipole interaction. However, if the grain has a high level of iron inclusions (i.e., $N_{\rm cl} > 10^2$), the collision rate by the Barnett dipole interaction could be enhanced with $(b_{\max}/a)^{2} > 100$, even in a dense environment with $\gamma U/n_{1} \sim 0.1$ (see in the bottom panels of Figure \ref{fig:bmax_a_SPM_vther_T100}).

\begin{figure*}
    \centering
    \includegraphics[width = 1\textwidth]{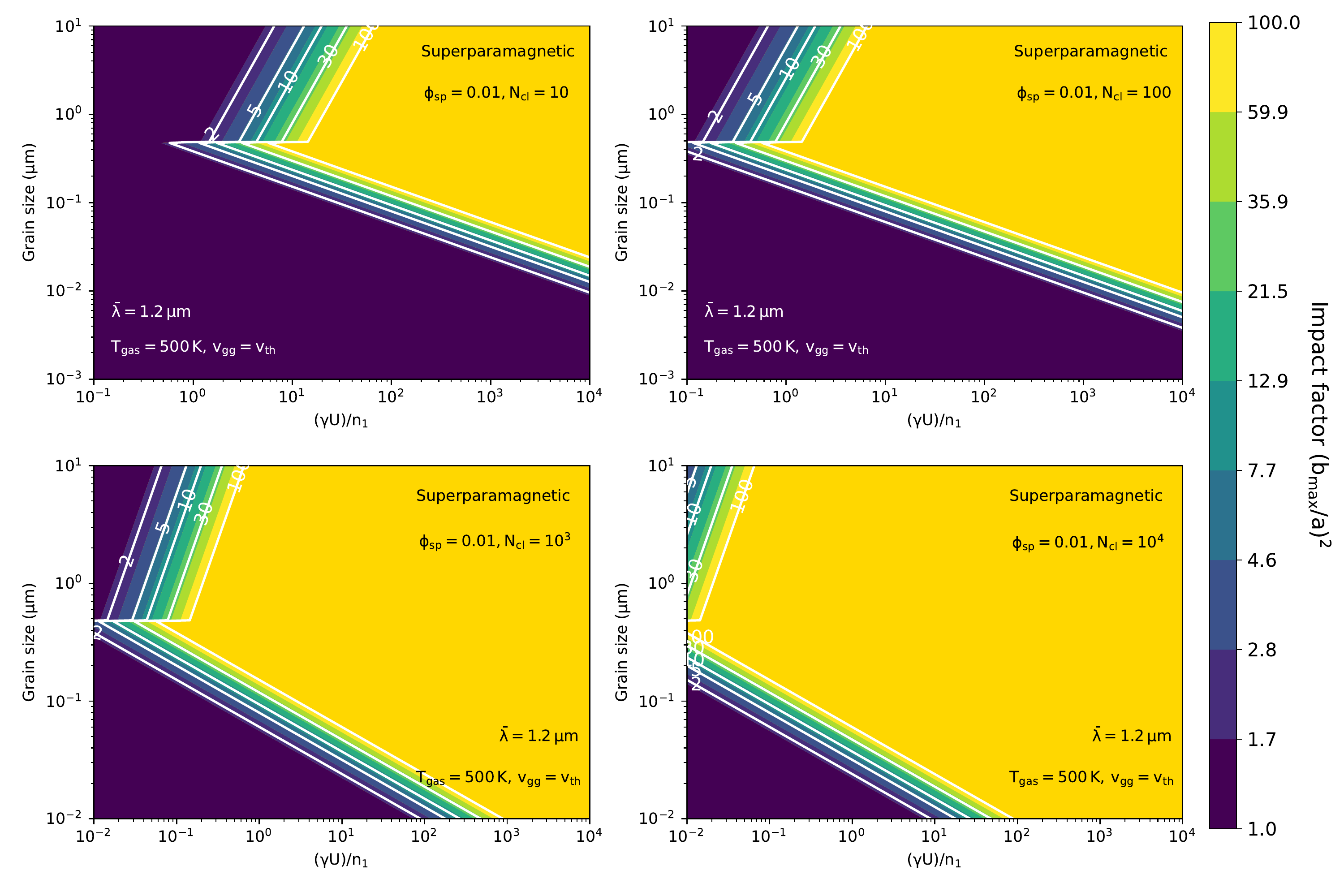}
    \caption{Same as Figure \ref{fig:bmax_a_SPM_vther_T100} but with $\rm T_{\rm gas} = 500\K$. The cross-section $(b_{\rm max}/a)^2$ tends to decrease due to the efficient gas collision damping at higher gas temperature environments.}
    \label{fig:bmax_a_SPM_vther_T500}
\end{figure*}

Figure \ref{fig:bmax_a_SPM_vther_T500} shows the similar results as in Figure \ref{fig:bmax_a_SPM_vther_T100} but for grains in a warmer region with $T_{\rm gas}=500\K$. The Barnett dipole interaction is still efficient, but less efficient than the case of $T_{\rm gas}=100\K$ due to both faster relative grain motion and the damping by gas collisions. For instance, the collision rate $(b_{\max}/a)^{2}$ decreases to $\sim 2$ for sub-micron SPM grains with $a = 0.1\,\rm\mu m$ and $N_{\rm cl} = 10$.

\begin{figure*}
    \centering
    \includegraphics[width = 1\textwidth]{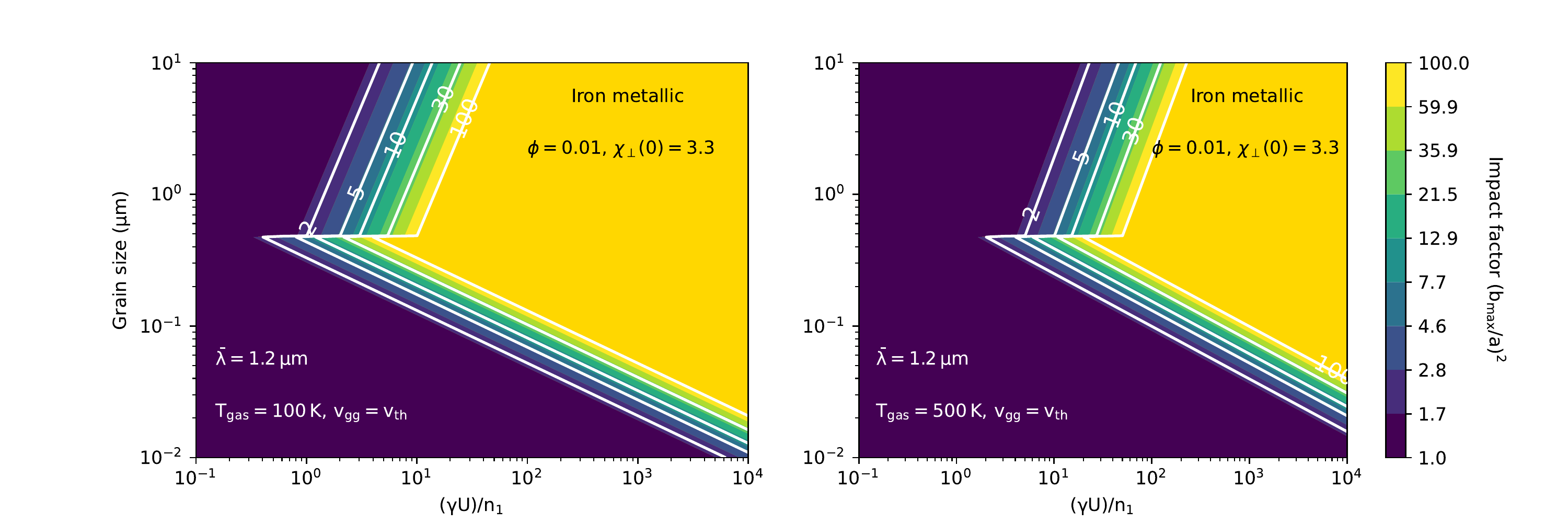}
    \caption{The enhanced collision rate, $(b_{\rm max}/a)^2$, of iron metallic grains moving at a thermal speed $v_{\rm th}$ with $T_{\rm gas} = 100\K$ (left panel) and $T_{\rm gas} = 500\K$ (right panel). The Barnett dipole interaction is less efficient due to lower magnetic susceptibility produced by single-domain iron metallic inclusions, compared with the case of SPM grains in Figure \ref{fig:bmax_a_SPM_vther_T100} and \ref{fig:bmax_a_SPM_vther_T500}.}
    \label{fig:bmax_a_ferro_vthermal}
\end{figure*}

Figure \ref{fig:bmax_a_ferro_vthermal} shows the numerical calculation of the enhanced collision rate $(b_{\rm max}/a)^2$ for iron metallic grains. As seen, the Barnett dipole-dipole interaction can also enhance the collision rate of two grains containing single-domain iron metallic irons for the local conditions of $\gamma U/n_1 >1$. However, in comparison with the SPM grains in Figures \ref{fig:bmax_a_SPM_vther_T100} and \ref{fig:bmax_a_SPM_vther_T500}, the Barnett dipole interaction is less efficient due to the lower magnetic susceptibility (see Equation \ref{eq:muBar_fer}). 

\section{Discussion}\label{sec:discuss}
Here we discuss the implications of the Barnett dipole-dipole interaction for grain growth in potential astrophysical environments.

\subsection{The role of the Barnett dipole-dipole and grain growth and destruction}
The grain-grain collision timescale in the Barnett effect is given by
\bea
t_{\rm coll}&=&\frac{1}{R_{\rm coll}}=\left(\frac{b_{\rm max}}{a}\right)^{-2}t_{\rm coll}^{0}\nonumber\\
&=&\left(\frac{b_{\rm max}}{a}\right)^{-2}\left(\frac{1}{n_{\rm gr}\pi a^{2}v_{\rm gg}}\right),\label{eq:tgg}
\ena
where $t_{\rm coll}^{0}$ is the grain-grain collision time without the Barnett effect and $n_{\rm gr}$ is the number density of dust grains.

Equations (\ref{eq:ba_RAT_SPM1}) and (\ref{eq:ba_RAT_SPM2}) reveal that the enhancement in the grain collision rate by Barnett dipole-dipole effect depends on dust magnetic properties ($\phi_{\rm sp},N_{cl}$), grain sizes, the radiation field and gas properties, $\gamma U/n_{1}T_{1}$, and the grain relative velocity. For a given local radiation field and gas density, the rate of grain growth sensitively depends on the grain's magnetic properties. Grains with a higher abundance of iron (larger $\phi_{\rm sp}$ and $N_{cl}$) have a higher collision rate due to the stronger Barnett dipole interaction. As a result, magnetic grains can collide and grow faster than carbonaceous or silicate grains without iron clusters. This process leads to the accelerating growth of superparamagnetic grains, which eventually form the iron-rich.

\subsection{Grain coagulation in the envelope of AGB and RSG stars}
Collisions between grains in the envelope are the primary process for grain growth. Note that iron grains are expected to form \citep{Jones.1990,Kemper.2002} or incorporated gradually in silicate grains \citep{Hofner.2022} in the inner envelope of O-rich Asymptotic Giant Branch (AGB) stars. Given the strong radiation field in the envelope, grains should spin rapidly. For the circumstellar envelope of AGB stars with the magnetic field strength of $B\sim 10(100\AU/r)^{-2}$ mG for a dipole magnetic field (see \citealt{Vlemmings.2018}), and the hydrogen density profile of
\bea
n_{\H}&\simeq& 4.5\times 10^{5}\left(\frac{\dot{M}}{ 10^{-5}M_{\odot}\,{\rm yr}^{-1}}\right)\left(\frac{10\,{\rm  km~s}^{-1}}{v_{\rm exp}}\right)\nonumber\\
&\times&\left(\frac{100\AU}{r}\right)^{2}
 \,{\rm cm}^{-3},~~~
 \ena
 where $\dot{M}$ is the mass-loss rate, and $v_{\rm exp}$ is the expansion velocity of the envelope (\citealt{2020ApJ...893..138T}). 
 
Using Equation (\ref{eq:delta_m}), one can see that magnetic grains are aligned perfectly by MRAT for $N_{\rm cl,4}\phi_{\rm sp,-2}>10^{-3}$ or $N_{\rm cl}\phi_{\rm sp,-2}>10$. Therefore, grains with iron inclusions can have perfect alignment by MRAT.

\begin{figure*}
    \centering
    \includegraphics[width = 1\textwidth]{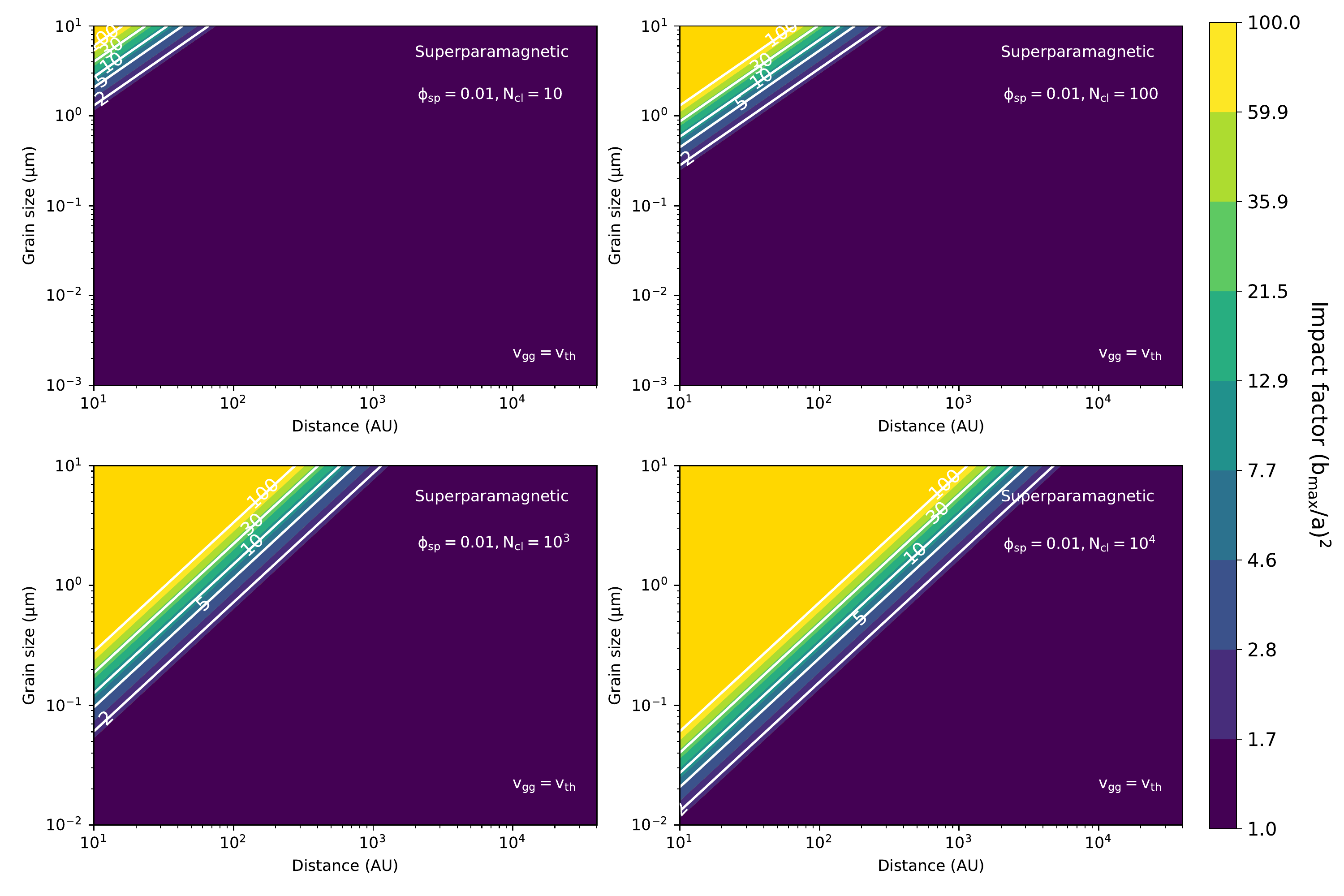}
    \caption{The enhancement of collision rate by dipole-dipole interaction induced by ambient magnetic fields with respect to the envelope distance, adopting magnetic field strength and gas properties of O-rich AGB stars (i.e., IK Tau, see in \citealt{Vlemmings.2018} and \citealt{2020ApJ...893..138T}). With the impact of strong magnetic field ($1\,\rm G - 100$ mG), large grains could be efficiently grown by dipole interaction in the inner regions of AGB envelopes.}
    \label{fig:bmax_a_AGB_mag}
\end{figure*}

The interaction of induced dipoles by ambient magnetic fields would become more efficient for large magnetic fields. Figure \ref{fig:bmax_a_AGB_mag} shows these effects on grain collision in the envelope of a particular AGB star - IK Tau varying from 10 AU to 40,000 AU, with a stellar magnetic field (1 G - 100 mG) and a gas density profile of $n_{\rm H}  \thicksim 1/r^{-2}$ (see more in  \citealt{Vlemmings.2018} and \citealt{2020ApJ...893..138T}). One can see that, as a result of strong magnetic fields in the inner regions of the AGB envelope, the dipole interaction is significantly effective for large SPM grains $a > 0.1\,\mu m$. And the enhancement of grain collision rate extends from the inner to the outer envelope with increasing $N_{\rm cl}$ due to higher magnetic susceptibility (bottom panels of Figure \ref{fig:bmax_a_AGB_mag}). 




In the envelope of red supergiant stars (RSGs), the typical density and radiation field is $\gamma U/n_{1}\sim 10-100$ (see \citealt{Truong.2022}). Using the results from Figure \ref{fig:bmax_a_SPM_vther_T100} and \ref{fig:bmax_a_SPM_vther_T500}, one can see that the cross-section is significantly enhanced for metallic grains or grains with iron inclusions. As a result, the growth of metallic grains or grains with iron inclusions. The accelerated dust coagulation can rapidly form large grains inside dust clumps in RSG envelopes.

\subsection{Grain growth and destruction in PDRs}
In PDRs, depending on the distance to the star, the typical density and radiation field is $U/n_{1}\sim 1-10$. For the Orion Bar, one has $U=4\times 10^{4}$ and $n_{\H}\sim 5\times 10^{5}\cm^{-3}$ (see Tielens), which corresponds to $U/n_{1}=2.7$. Using the results from Figure \ref{fig:bmax_a_SPM_vther_T100} and \ref{fig:bmax_a_SPM_vther_T500} one can see that the cross-section is significantly enhanced. As a result, grain growth can occur faster due to enhanced grain collisions. 


\subsection{Grain growth in protostellar environments}
In hot cores/corinos surrounding the Class 0/I protostar, grains are subject to strong protostellar radiation. Therefore, the Barnett dipole-dipole interaction can be efficient in enhancing grain collisions and grain growth. A similar effect can occur in the inner envelope of Class 0 protostars. A detailed modeling for grain growth will be presented in our follow-up paper. For the conditions of low-mass protostellar cores, one can have $U\sim 10^{3}-10^{6}$ and $n_{\H}\sim 10^{7}$ (see \citealt{Giang.2023}). In this case, one has $U/n_{1}\sim 10^{-3}-1$. From Figure \ref{fig:bmax_a_SPM_vhydro_T100}, one can see that the Barnett dipole effect is important in the hot cores/corinos for grains with large iron inclusions of $N_{\rm cl}=10^{3}-10^{4}$.





\subsection{Effect of grain acceleration by gas turbulence}
In the presence of interstellar turbulence, grains can be accelerated by hydrodynamic turbulence (\citealt{1985prpl.conf..621D}; \citealt{2002ApJ...566L.105L}). In this case, the grain velocity increases with the grain size because they are affected by larger turbulent eddies that have higher turbulent energy. For grains in molecular clouds, the turbulence driving scale is the Jean length. Then, the relative grain velocity by hydrodynamic turbulence is given by (see \citealt{2013MNRAS.434L..70H})
\bea
v_{\rm turb}(a)\simeq 1.86\times 10^{3}\hat{\rho}^{1/2}T_{2}^{1/4}a_{-5}^{1/2}n_{5}^{-1/4}\cm\s^{-1},\label{eq:vgr_turb}
\ena
where $n_{5}=n_{\H}/10^{5}\cm^{-3}$.

\begin{figure*}
    \centering
    \includegraphics[width = 1\textwidth]{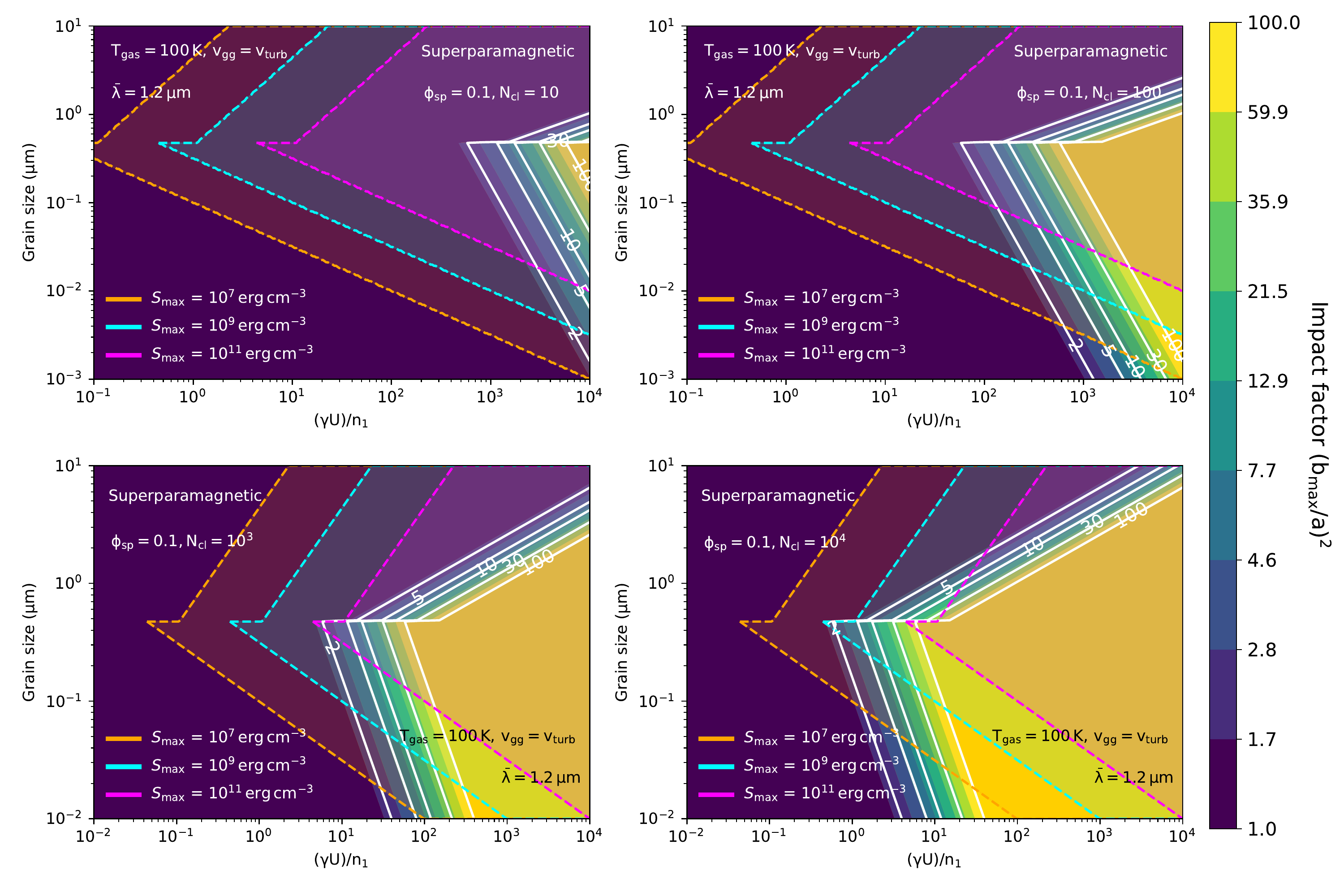}
    \caption{Enhanced collision rate $(b_{\rm max}/a)^2$ of SPM grains driven by hydrodynamic turbulence (i.e., $v_{\rm gg} = v_{\rm turb}$). Small grains $a < 1\,\rm\mu m$ obtain lower speed produced by small eddies (i.e., lower turbulence energy), resulting in the enhancement of grain growth by Barnett dipole interaction.}
    \label{fig:bmax_a_SPM_vhydro_T100}
\end{figure*}

We calculate the enhancement in collision rates assuming $v_{\rm gg}=v_{\rm turb}$ for the different physical parameters. Figure \ref{fig:bmax_a_SPM_vhydro_T100} shows the increase in the collision rate of SPM grains under the effect of gas turbulence, assuming a value of $\phi_{\rm sp}=0.1$ and $T_{\gas}=100\K$. The impact of Barnett dipole interaction on grain growth is still significant when the grain is exposed to strong radiation fields (i.e., $\gamma U/n_1 > 10$), and associated with the increasing levels of iron inclusions. With the presence of hydrodynamic turbulence, sub-micron grains ($a < 0.1\,\rm\mu m$) are driven by small-scale turbulent eddies with lower turbulence energy, and thus, they are moving at lower relative velocity. They could easily collide with each other by the Barnett dipole interaction, resulting in the enhancement of collision cross-section with $(b_{\rm max}/a)^2 > 10$, compared with the case of large grains $a > 1\,\rm\mu m$ with $(b_{\rm max}/a)^2 < 10$. The results raise the importance of gas turbulence in grain growth by magnetic dipole interaction and imply the possibility of the growth from nanoparticles (i.e., $a < 100\,\rm\AA$) and sub-micron grains (i.e., $a < 0.1\,\rm\mu m$) to very large grains (i.e., $a > 10\,\rm\mu m$) in hydrodynamic turbulence environments such as PDRs and protostellar cores. 

Note that charged grains can be accelerated by magnetohydrodynamic turbulence (\citealt{2003ApJ...592L..33Y,Yan:2004ko}; \citealt{Hoang:2012cx}), which is not considered here.

\subsection{Effects of Radiative Torque Disruption (RAT-D)}
As experiencing suprathermal rotation by RATs, the grain develops centrifugal stress $S = \rho \Omega^2 a^2/4$ on grain materials. If the centrifugal stress exceeds the binding energy of grains, they are spontaneously fragmented into many smaller species, which is called Radiative Torque Disruption (RAT-D) (see more in \cite{Hoang:2019da}).

The critical angular velocity at which the grain is disrupted is calculated as 
\bea
\Omega_{\rm disr} &=& \frac{2}{a}\,\left(\frac{S_{\rm max}}{\rho}\right)^{1/2}\nonumber\\
&\simeq& \frac{3.65\times10^{8}}{a_{-5}}\hat{\rho}^{-1/2}S_{\rm max,7}^{1/2}\,\rm rad\,s^{-1},
\ena
where $S_{\rm max,7}=S_{\rm max}/(10^7\erg\cm^{-3})$ is the maximum tensile strength of grains (\citealt{Hoang:2019da}), which is characterized by the grain internal structure.  Porous/composite grains have low values of $S_{\rm max} = 10^{6} - 10^8\,\rm erg\,\rm cm^{-3}$, while compact grains have higher values as $S_{\rm max} = 10^{9} - 10^{10}\,\rm erg\,\rm cm^{-3}$.

Dust grains undergo rotational disruption when the grain angular velocity spun-up by RATs, $\Omega_{\rm RAT}$, exceeds the critical velocity of $\Omega_{\rm disr}$. The minimum size $a_{\rm disr}$ that grains can be disrupted by RATs is defined as
\bea
a_{\rm disr} &=& \left(\frac{0.8n_{\H}\sqrt{2\pi m_{\H}kT{\rm gas}}}{\gamma u_{\rm rad}\bar{\lambda}^{-2}}\right)^{1/2} \left(\frac{S_{\rm max}}{\rho}\right)^{1/4}(1 + F_{\rm IR})^{1/2}\nonumber\\
&\simeq&
53.7\,\left(\frac{\gamma U}{n_1 T_1}\right)^{1/2}\left(\frac{\bar{\lambda}}{\rm 1.2\,\mu m}\right)\hat{\rho}^{-1/4}S_{\rm max,7}^{1/4}\nonumber\\
&\times&(1 + F_{\rm IR})^{1/2}\,\rm \mu m,
\ena
and the maximum grain size still being disrupted by RAT-D is 
\bea
a_{\rm disr, max} &=& \frac{\gamma u_{\rm rad}\bar{\lambda}}{16n_{\H}\sqrt{2\pi m_{\H}kT{\rm gas}}} \left(\frac{S_{\rm max}}{\rho}\right)^{-1/2}(1 + F_{\rm IR})^{-1}\nonumber\\
&\simeq&
30\,\left(\frac{\gamma U}{n_1 T_1}\right)\left(\frac{\bar{\lambda}}{\rm 1.2\,\mu m}\right)\hat{\rho}^{1/2}S_{\rm max,7}^{-1/2}\nonumber\\
&\times&(1 + F_{\rm IR})^{-1}\,\rm \mu m,
\ena
which depends on the radiation field $\gamma U$, the local gas properties $n_1T_1$ and grain internal structure $S_{\rm max}$. Under the effects of RAT-D, the grain growth by Barnett dipole-dipole interaction is efficiently constrained in the local environment with strong radiation strength $U$ and less gas density $n_{\H}$.

In Figure \ref{fig:bmax_a_SPM_vhydro_T100}, the disruption is efficient for disrupting large grains, particularly for porous grains with $S_{\rm max} < 10^{9}\erg\cm^{-3}$ even in the environments of weaker radiation field strength (i.e., $\gamma U/n_1 < 1$). For grains with small iron clusters, RAT-D is more effective than the dipole-dipole interaction. However, for larger iron clusters, the dipole-dipole interaction is effective for small grains of $a<0.1-0.5\mum$. Beyond this size, grains are disrupted by RAT-D for $S_{\rm max} > 10^{9}\erg\cm^{-3}$. 


\section{Summary}\label{sec:summary}
We study the effect of Barnett magnetic dipole-dipole interaction on grain collisions and our main results are summarized as follows.

\begin{enumerate}

\item Rapidly spinning magnetic grains spun up by RATs acquire large magnetic moments due to the Barnett effect. The interaction between resulting Barnett magnetic moments can significantly enhance the collision cross-section of spinning grains and the grain collision rate.

\item The enhanced grain collision rate by the Barnett magnetic moments can be important for grain growth and destruction in the radiation-dominated regions such as the circumstellar envelope of evolved stars, PDRs, and central protostellar cores. 

\item In strong radiation fields, the rate of grain growth is significantly enhanced by Barnett dipole-dipole interaction. On the other hand, the rotational disruption by RATs (RAT-D mechanism) can efficiently destroy large grains. Thus, the combined effect of the Barnett dipole-dipole and RAT-D can speed up the process of grain coagulation and destruction.

\item Our results suggest the importance of the dust magnetic properties and the radiation field on grain growth and disruption, which must be taken into account for accurate modeling of grain evolution.

\end{enumerate}

\acknowledgments
T.H. acknowledges the support by the National Research Foundation of Korea (NRF) grant funded by the Korea government (MSIT) (2019R1A2C1087045). This work was partly supported by a grant from the Simons Foundation to IFIRSE, ICISE (916424, N.H.). We would like to thank the ICISE staffs for their enthusiastic support.


\bibliography{ms.bbl}
\end{document}